\documentstyle[prl,aps,twocolumn,epsfig]{revtex}
\begin{document}

\title{Vorticity Statistics In The Two-Dimensional Enstrophy Cascade}
\author{  J\'er\^ome Paret, Marie-Caroline Jullien and Patrick 
Tabeling}
\address{Laboratoire de Physique Statistique, Ecole Normale 
Sup\'erieure,\\
24 rue Lhomond, 75231 Paris (France)}
\date{\today}
\maketitle
\begin{abstract}
We report the first extensive experimental observation of the 
two-dimensional enstrophy cascade, along with the determination of the 
high order vorticity statistics.  The energy spectra we obtain are 
remarkably close to the Kraichnan Batchelor expectation.  The 
distributions of the vorticity increments, in the inertial range, 
deviate only little from gaussianity and the corresponding structure 
functions exponents are indistinguishable from zero.  It is thus shown 
that there is no sizeable small scale intermittency in the enstrophy 
cascade, in agreement with recent theoretical analyses. 

\end{abstract}
\pacs{47.27.Gs}

 The enstrophy cascade is one of the most important processes in 
 two-dimensional turbulence, and its investigation, at 
 fundamental level, provides cornerstones for the analysis of 
 atmosphere dynamics. The existence of this cascade was first 
 conjectured by Kraichnan \cite{kra1}, and later by Batchelor 
 \cite{bat}.  Both of them proposed that in two-dimensional 
 turbulence, enstrophy injected at a prescribed scale is 
 dissipated at smaller scales, undergoing a cascading process at 
 constant enstrophy transfer rate $\eta$ ; this led to predicting 
 a $k^{-3}$ spectrum for the energy, in a range of scales extending 
 from the injection to the dissipative scale.  Later, logarithmic 
 corrections have been incorporated in the analysis to ensure 
 constancy of the enstrophy transfer rate \cite{kra2}.  The 
 advent of large computers revealed surprising deviations from 
 the classical expectation, especially in decaying systems 
 \cite{san,bra,leg,okh}.  It was soon realized that in 
 two-dimensional systems, long live coherent structures inhibit 
 the cascade locally and therefore the self similarity of the 
 process, assumed to fully apply in the classical approach, is 
 broken.  Expressions like "laminar drops in a turbulent 
 background" were coined to illustrate the role of coherent 
 structures in the problem \cite{san}.  Along with the 
 observations of unexpected exponents, models, emphasizing on the 
 role of particular vortical structures \cite{saf,mof}, or based 
 on conformal theory \cite{pol}, suggested non classical values.  
 In the recent period however, high resolution simulations 
 \cite{her,bor,mal,oet} underlined that, provided long live 
 coherent structures are disrupted, classical behaviour holds ; 
 furthermore, theoretical studies \cite{eyi,fal} 
 suggested the absence of small scale intermittency, placing the 
 direct enstrophy cascade in a position strikingly different from the 
 three-dimensional energy cascade.  The recent soap film experiments, 
 developing single point measurements of the velocity field 
 \cite{gha,kel,mar,rut}, obtained spectral exponents consistent 
 with these views. 
 
 Nonetheless, investigating small scale intermittency in this problem 
 requires measuring the statistics of quantities such as the vorticity 
 increments, which has not been done yet, neither in physical nor in 
 numerical experiments.  Efforts in this direction were made in the 
 numerical study of Borue \cite{bor}, but difficulties arose to obtain 
 converged results.  An analysis of the enstrophy fluxes in the 
 numerical experiment of Babiano et al \cite{bab} led the authors to 
 underlining the presence of weak intermittency in the enstrophy 
 cascade ; thus, although the theory on the problem is at a well 
 advanced stage (at least compared to three-dimensional situation), it 
 is not yet known, even in situations where self similarity fully 
 holds, to what extent classical theory, based on mean field arguments 
 "a la Kolmogorov", can be applied to describe the enstrophy cascade.  
 In the physical experiment we present here, we have extensively 
 measured the statistics of vorticity increments, in a situation where 
 coherent structures have been disrupted.  We could show, for the 
 first time, that in the enstrophy cascade, the deviation from 
 gausiannity, for the small scale statistics of the vorticity field, 
 are moderate, and -more importantly- scale independant ; the 
 corresponding structure function exponents are indistinguishable from 
 zero, so that intermittency is absent from the process, in agreement 
 with the theoretical analysis of \cite{fal}.  This 
 observation, made on a physical system perhaps brings the problem, 
 more firmly, within the reach of a theoretical understanding, a 
 situation rare in the field.
 
The experimental set-up has been described in a series of papers 
\cite{han,par1,par2}.  It appears that the system we use is a 
formidable tool for investigating fundamental issues of 
two-dimensional turbulence.  It provides reliable data on quantities 
reputed hard to measure.  We believe this is an interesting situation, 
since it would be unpleasant to elaborate a rationale for 2D 
turbulence, solely on virtual inputs .  Briefly speaking, the flow is 
generated in a square PVC cell, 15 cm $\times$ 15 cm.  The bottom of 
the cell is made of a thin (1 mm thick) glass plate, below which 
permanent magnets, 5x8x4 mm in size, and delivering a magnetic field, 
of maximum strength 0.3 T, are placed.  In order to ensure 
two-dimensionality \cite{par3}, the cell is filled with two layers of 
NaCl solutions, 2.5 mm thick, with different densities, placed in a 
stable configuration, i.e.  the heavier underlying the lighter.  Under 
typical operating conditions, the stratification remains unaltered for 
periods of times extending up to 10mn.  The interaction of an 
electrical current driven across the cell with the magnetic field 
produces local stirring forces.  The flow is visualized by using 
clusters of 2 $\mu$m in size latex particles, placed at the free 
surface, and the velocity fields ${\bf v}({\bf x},t)$ are determined 
using particle image velocimetry technique, implemented on $64 \times 
64$ grids.  In such experiments, the dissipative scale for the 
enstrophy cascade - defined by $l_{d}=\eta^{1/6}{\nu^{1/2}}$ (where 
$\nu$ is the kinematic viscosity of the working fluid) - is on the 
order of 1 mm ; it is thus unresolved.  Moreover, $l_{d}$ lying below 
the layer thickness, it is reasonable to consider that the way how 
enstrophy is dissipated in our system is not purely two-dimensional.  
Concerning now measurement accuracy, we estimate, from the measurement 
of local divergence, that the accuracy on the velocity is a few 
percent and that on the vorticity is 10\%.

In the experiments we describe here, magnets are arranged into four 
triangular aggregates of roughly one hundred units, having the same 
magnetic orientation, as shown schematically in Fig \ref{forcing}.  By 
doing so, the electromagnetic forcing is defined on large scale, and 
its spatial structure does not favour any particular permanent 
pattern.

The electrical current is unsteady : it is a non periodic, zero mean, 
square waveform, of amplitude equal to 0.75 A (see Fig \ref{forcing}).  
The corresponding Reynolds number - defined as the square of the ratio 
of the forcing to the dissipative scale - is on the order of $10^{3}$ 
; this estimate is one order of magnitude above the largest simulation 
performed on the subject, using normal viscosity (See \cite{bor}).  In 
the statistically steady state, the instantaneous flow pattern 
consists of transient recirculations of sizes comparable to one fourth 
of the box size.  The formation of permanent large scale structures, 
which might tend to break the self similarity of the process, seems 
disrupted by our particular forcing. 

 The instantaneous vorticity 
field in the statistically stationary state is shown in Fig 
\ref{champvort}.  We see elongated structures, in form of filaments or 
ribbons, some of them extending across a large fraction of the cell.  
At variance with the decaying regimes, and consistently with the above 
discussion, we have not seen any long live vorticity concentration, 
i.e persisting more than a few seconds.  This is further confirmed by 
a measurement of the flatness of the vorticity distribution, a 
diagnostics previously introduced by \cite{mal} and which is found 
slightly above the gaussian value in our case.  The presence of 
coherent structures would have been associated to much larger values 
of this quantity.  The isotropy of the vorticity field is not obvious 
from the inspection of a single realisation, such as the one of Fig 
\ref{champvort} ; nonetheless, as will be shown later, the overall 
anisotropy level, obtained after statistical averaging, turns out to 
be reasonably small.

 The spectrum of the velocity field, averaged over 200 realisations, in 
 the statistically steady state, is shown in Fig \ref{spec}.  The 
 forcing wave number $k_{f} \sim $ 0.6 cm$ ^{-1}$ corresponds to the 
 location of the maximum of the energy spectrum ; it is associated to 
 an injection scale $l_{f}=\frac{2 \pi}{k_{{f}}}$ estimated to 10 cm, 
 a value consistent with the size of our permanent magnets clusters.  
 The wave-number associated to the stratified fluid layer, may be 
 defined as $k_{l}=\frac{2 \pi}{b}\sim $ 12 cm$ ^{-1}$ (where b is the 
 fluid thickness).  This wave-number, together with the sampling 
 wave-number, which is 25 cm$ ^{-1}$, are well outside the region 
 of interest. Fig \ref{spec} shows that in the high wave number 
 region, i.e above  9 cm$ ^{-1}$, the spectrum is flat.  
 This region is dominated by white noise ; it reflects a limitation 
 in the PIV technique to resolve low velocity levels at small scales.

The interesting feature is that there exists a spectral band, lying 
 between $k_{f}$ and $k_{max}\sim $7 cm$ ^{-1}$ , uncontaminated by a 
 possible interaction with the layer wave-number, in which a power 
 law behaviour is observed.  The corresponding exponent is close to 
 -3, as shown on the compensated spectrum.  A direct measurement of 
 the exponent, performed by using least square fit in the scaling 
 region, leads to proposing the following formula for the spectrum :
 $$
 E(k) \sim  
k^{-3.0 \pm 0.2}
\label{spe}
$$
 The exponent we find is thus close to classical expectation.  There 
 is no steepening effect of the spectrum, which could be attributed, 
 as in decaying systems, to the presence of coherent structures.  
 Further analysis of the vorticity field shows homogeneity and 
 stationary, of the process.  
 Isotropy is also obtained, albeit only roughly, as shown in Fig 
 \ref{isotro} : to estimate the anisotropy level, we follow 
 circles, embedded in the inertial range, in the spectral plane of Fig 
 \ref{isotro}, and determine by how much the spectral energy departs 
 from a constant value along such circles.  This leads to an 
 anisotropy level on the order of 15 \% in
 the central region of the inertial range ; this is 
 acceptably low.  Determining 
 the Kraichnan Batchelor constant is a delicate issue, which entirely 
 relies on the measurement of the enstrophy pumping rate $\eta$.  The 
 constant we discuss here, called C', is defined by expressing the 
 energy spectrum in the form :
  $$
 E(k) = C' \eta^{2/3} k^{-3}
\label{spe2}
$$

 To measure C', we determine 
 the spectral enstropy transfer rate from below k to above k, 
 -  $\Lambda (k)$ - and search for a plateau, within the 
 scaling range of the energy spectrum.   
 $\Lambda (k)$ is found positive throughout this range, which 
 confirms the cascade is forward.  To determine $\eta$, we 
 further average out $\Lambda (k)$, between $k_{f}$ and $k_{max}$.  
 This procedure provides the following estimate for the Kraichnan Batchelor 
 constant C' :
 $$
C' \approx 1.4 \pm 0.3
\label{c'}
$$
This estimate agrees with that found in the high resolution study of 
Ref \cite{bor}, for which values ranging between $1.5$ and $1.7$ have been 
proposed.  We provide here the first experimental measurement ever 
achieved for this constant.

We now turn to the intermittency problem.  Fig \ref{pdf} shows a set 
of five distributions of the vorticity increments, obtained for 
different inertial scales, ranging between 2 and 9 cm.  As usual, in 
order to analyze shapes, the pdfs have been renormalized to impose 
their variance be equal to unity.  The shapes of the pdfs are not 
exactly the same, but there is no systematic trend with the scale 
across which the vorticity increment is determined.  Within 
experimental error, the distributions roughly collapse onto a single 
curve ; the tails of such an average distribution are broader than a 
gaussian curve, but here again, the deviations have a moderate 
amplitude and are scale independant.  It is thus difficult here, from 
the inspection of the distributions, to figure out the presence of 
intermittency in the enstrophy cascade.

The analysis of the structure functions of the vorticity, 
shown on Fig \ref{fstruc}, confirms this statement.  These
structure functions are defined by :
$$
 S_{p}(r) =<(\omega (\bf{x+r}) -\omega (\bf{x}))^{p}> 
\label{str}
$$
in which $\bf{x}$ and $\bf{r}$ are vectors, and r is the modulus of 
$\bf{r}$.  The brackets mean double averaging, both in space, 
throughout the plane domain, and in time,  between 20 
and 280 s.  We use here  $10^{5}$ data points to determine the 
structure functions ; this allows to determine up to twelfth order, 
because of the near gaussianity of the pdfs.  Fig \ref{fstruc} thus 
represents a series of vorticity structure functions $ S_{p}(r)$, 
obtained in such conditions, emphasizing on the inertial domain, i.e 
with r varying between 1 and 10 cm.  The structure functions 
are essentially flat, indicating the structure function exponents are 
close to zero.  The corresponding values of such exponents fall in the 
range -0.05, 0.15, for p varying between 2 and 10 ; this is 
indistinguishable from zero.  We thus obtain here a result fully 
compatible with the classical theory, for which the exponents are 
predicted to be exactly zero at all orders.  Surprisingly, we do not 
detect any logarithmic deviation, which would be compatible with the 
analysis of Refs \cite{fal}.  It is however not completely safe 
that this does not reflect a limitation in the measurements.

To summarize, we have performed, for the first time in a physical 
system, an extensive observation of the enstrophy cascade.  Previous 
experiments inferred its existence from the interpretation of $k^{-3}$ 
spectra.  We provide here a complete observation, along with a 
measurement of the Kraichnan Batchelor constant, and a determination 
of the high order vorticity statistics, a crucial quantity to measure 
for addressing the intermittency problem. We obtain that classical 
theory is strikingly successful ; there is no sizeable small scale 
intermittency, and the vorticity statistics departs only weakly from 
gausiannity, at all scales.  Because of these particular features, one 
may perhaps hope this problem 
be brought to theoretical understanding.  The role of coherent 
structures, long emphasized on, is indeed important and 
interesting, but should probably be considered as a separate issue.

 This work has been supported by Ecole 
Normale Sup\'erieure, Universit\'es Paris 6 et Paris 7, Centre 
National de la Recherche Scientifique, and by EEC Network Contract 
FMRX-CT98-0175.  The authors wish to thank G Falkovitch, V Lebedev, R 
Benzi for enlightening discussions concerning this study.

 \vskip1ex
\centerline{ \epsfysize=7cm \epsfbox{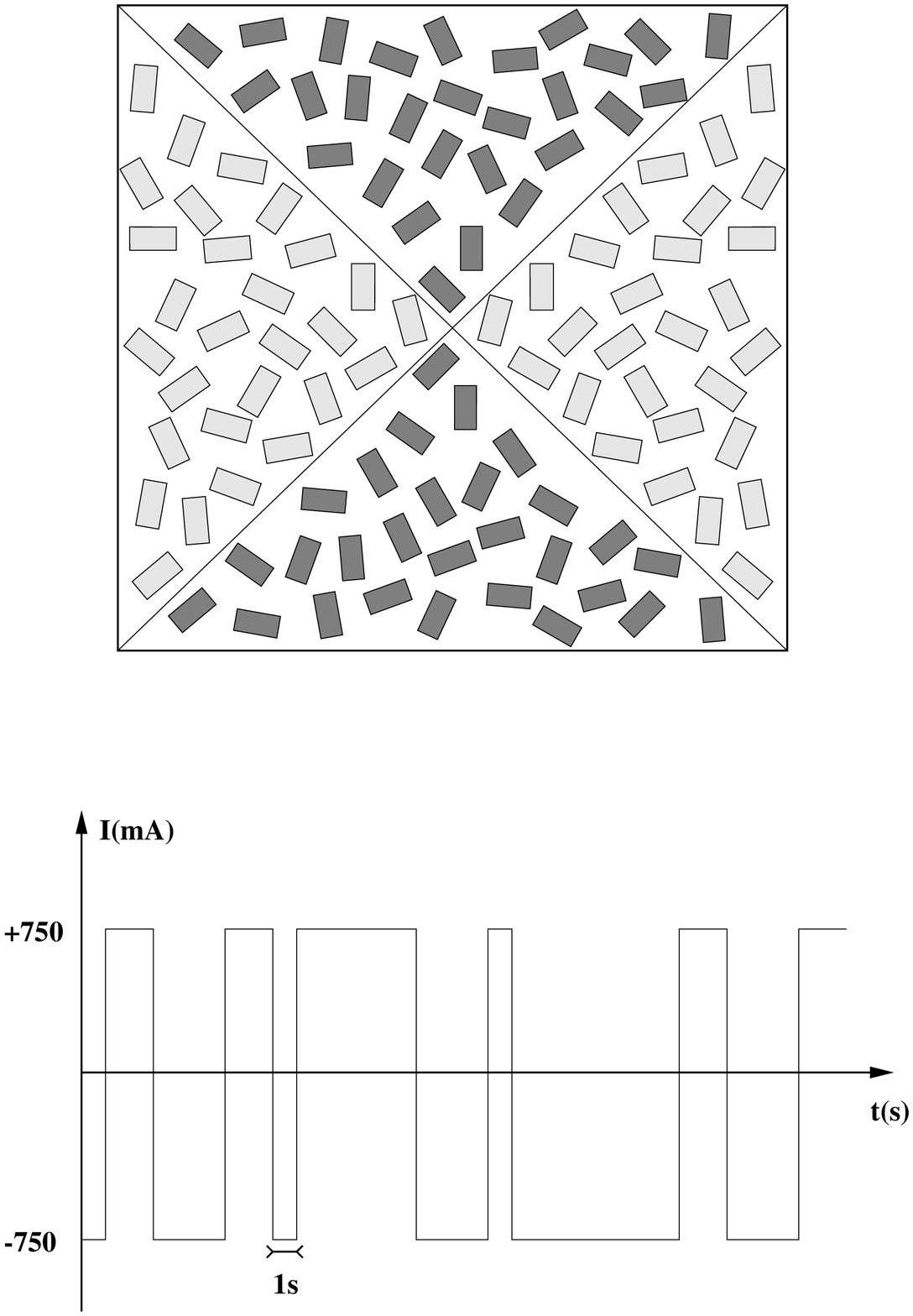} }
\begin{figure}
\caption{ A sketch of the organisation, in space, of the magnets (as seen 
from above) and the time dependence of the electrical current crossing 
 the cell. Black units have the same magnetic orientation, grey ones have 
 the opposite one. The averaged laps of time between two 
successive current switchs is 2.5 s}
\label{forcing}
\end{figure}

\vskip1ex
\centerline{
\epsfysize=6cm
\epsfbox{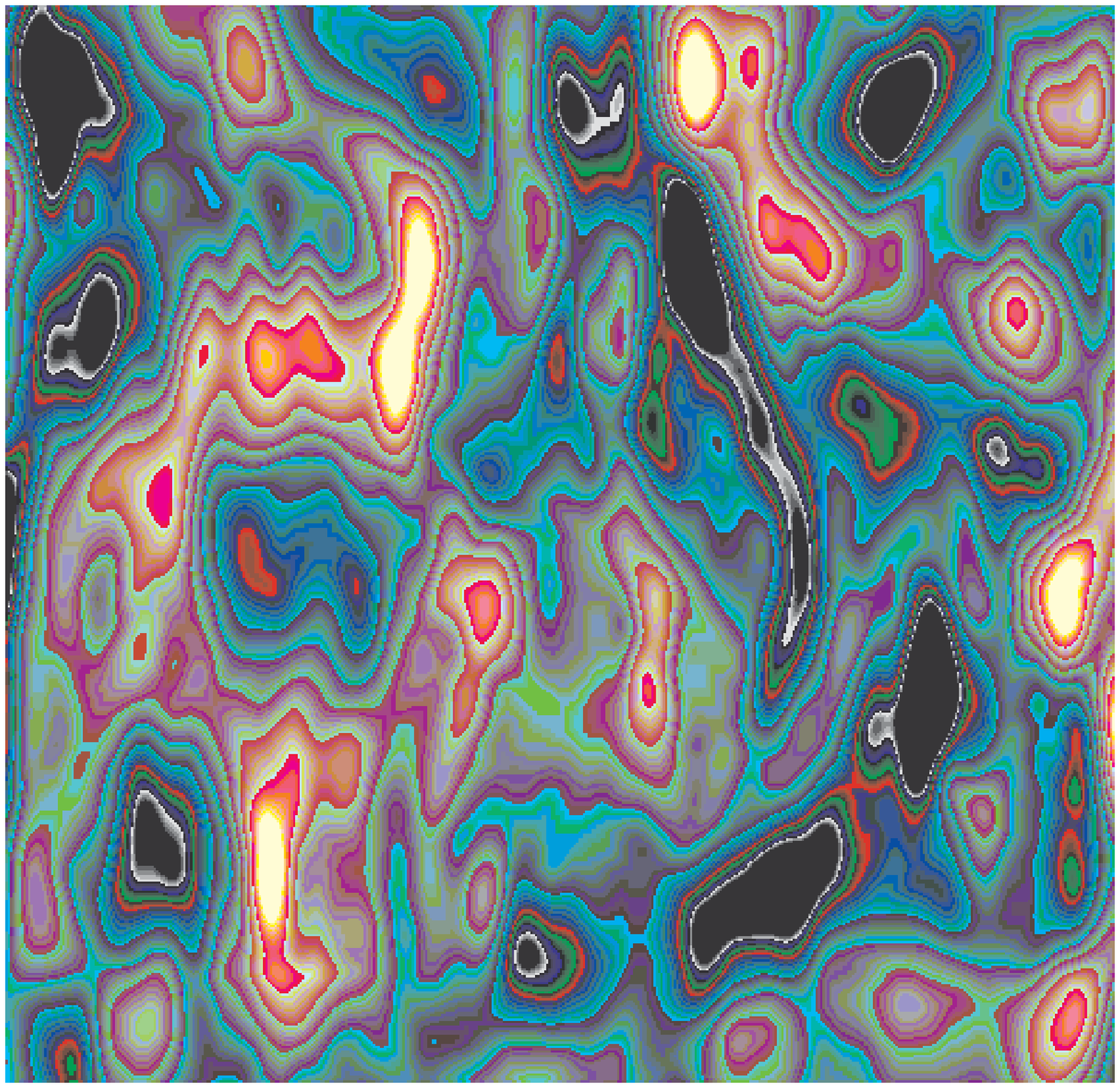}
}
\begin{figure}
\caption{ A particular realisation of the vorticity field, in the 
statistically steady state ; the grey scale is linear in the 
vorticity }
\label{champvort}
\end{figure}

 \vskip1ex
\centerline{
\epsfysize=6cm
\epsfbox{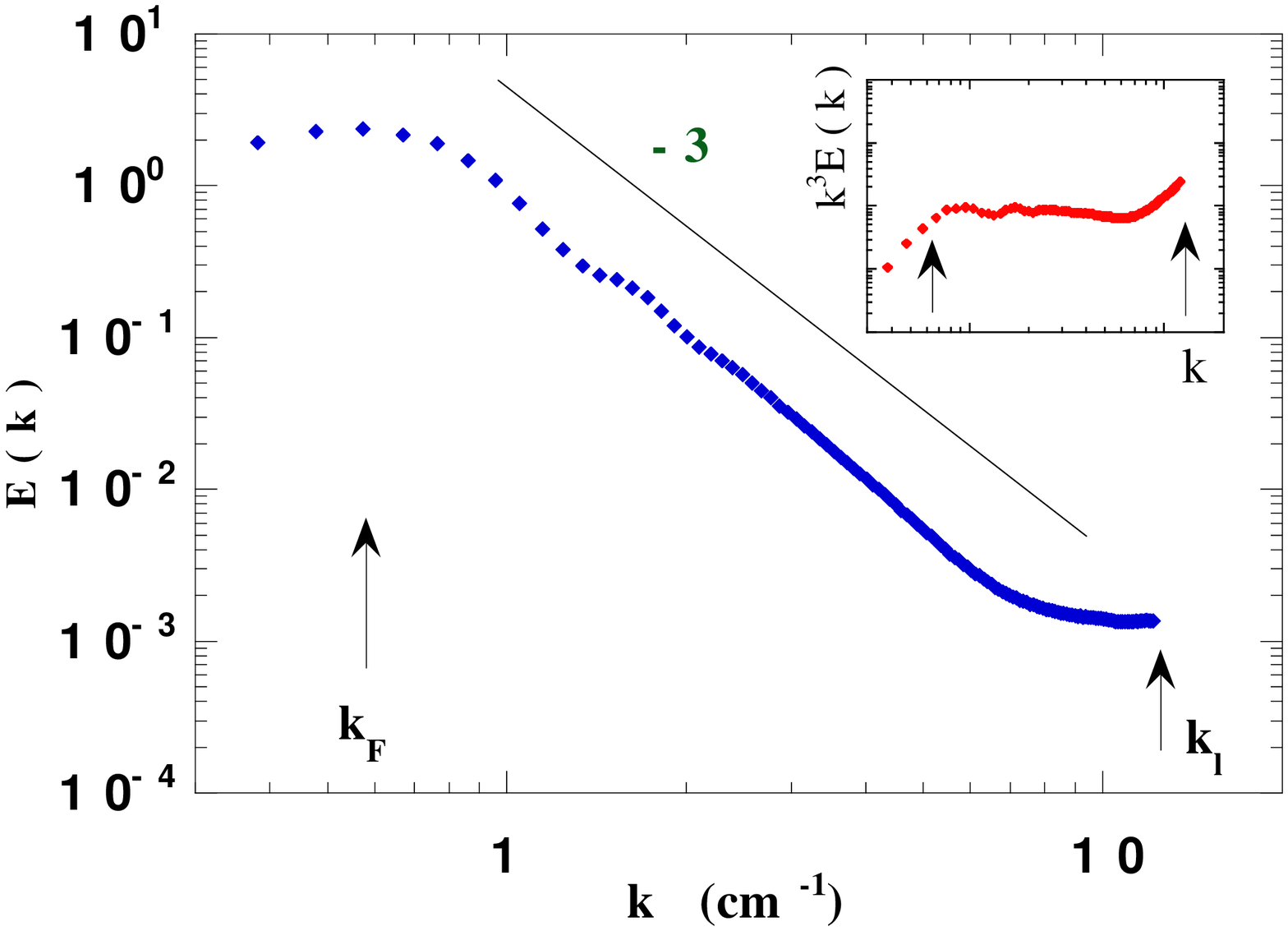}
}
\begin{figure}
\caption{Energy spectrum of the velocity field, averaged over 200 
realisations of the velocity field in the statistically stationary 
state ; the insert shows the same spectrum, multiplied by $k ^{3}$. }
\label{spec}
\end{figure}

 \vskip1ex
\centerline{
\epsfysize=4cm
\epsfbox{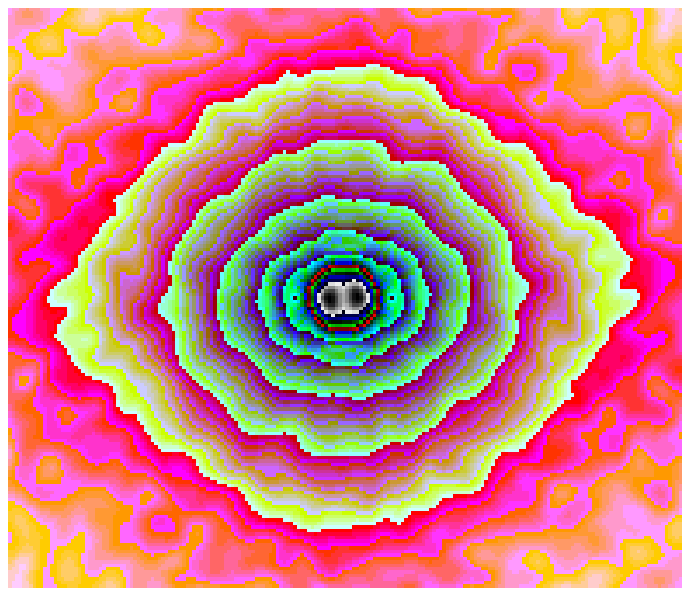}
}
\begin{figure}
\caption{ Two dimensional energy spectrum of the flow 
under investigation.  The two peaks at $k_{x}=\pm 0.6 cm^{-1}$ around 
the center (where x is the axis crossing these two peaks) 
signal the forcing.  The boundaries of the rectangle, along x axis 
corresponds to $k_{x} = \pm$ 12 cm$ ^{-1}$ }.
\label{isotro}
\end{figure}

 \vskip1ex
\centerline{
\epsfysize=5cm
\epsfbox{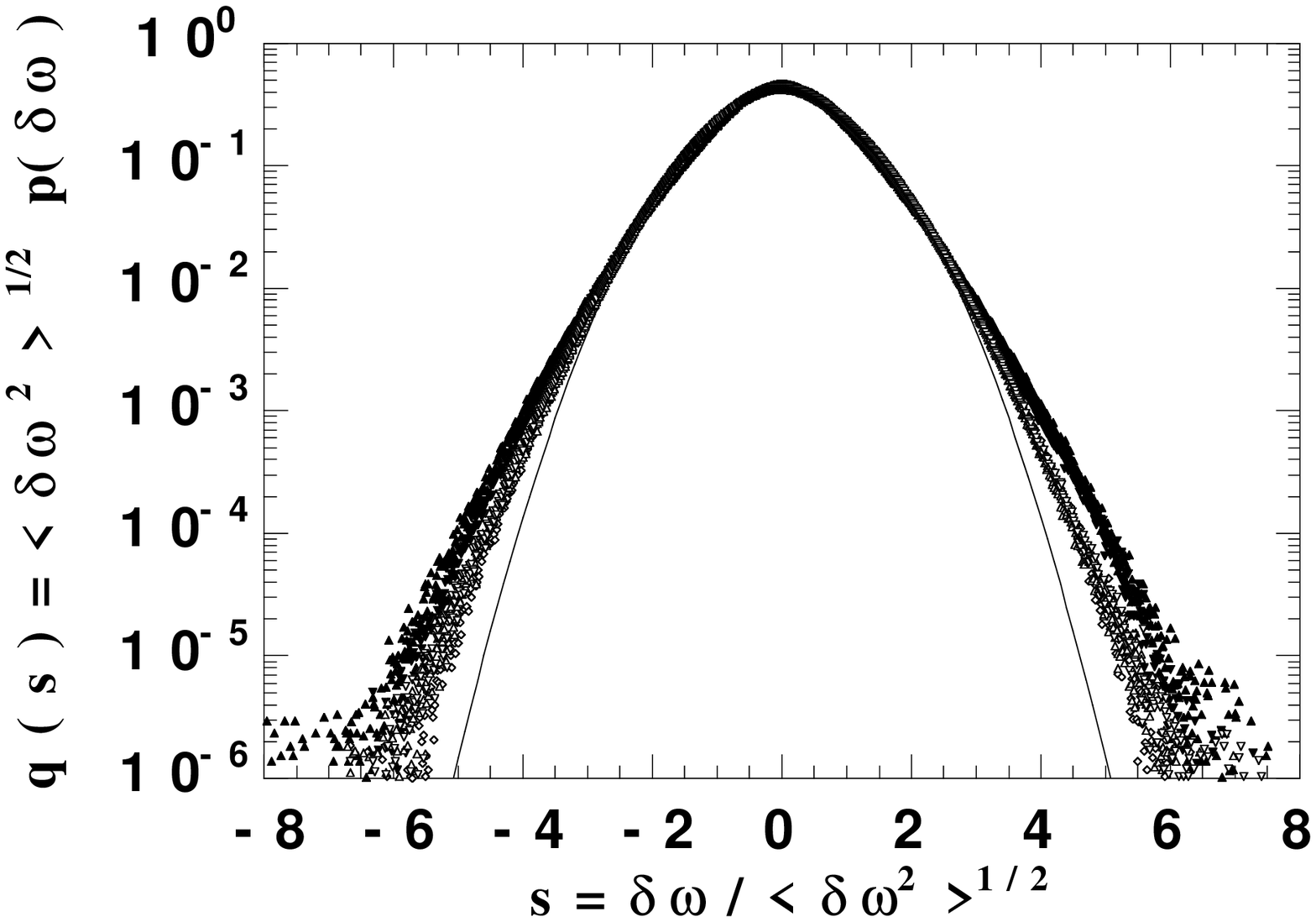}
}
\begin{figure}
\caption{Normalized distributions of vorticity increments, for five 
separations $r$ : 2, 3, 5, 7 and 9 cm.}
\label{pdf}
\end{figure}

 \vskip1ex
\centerline{
\epsfysize=5cm
\epsfbox{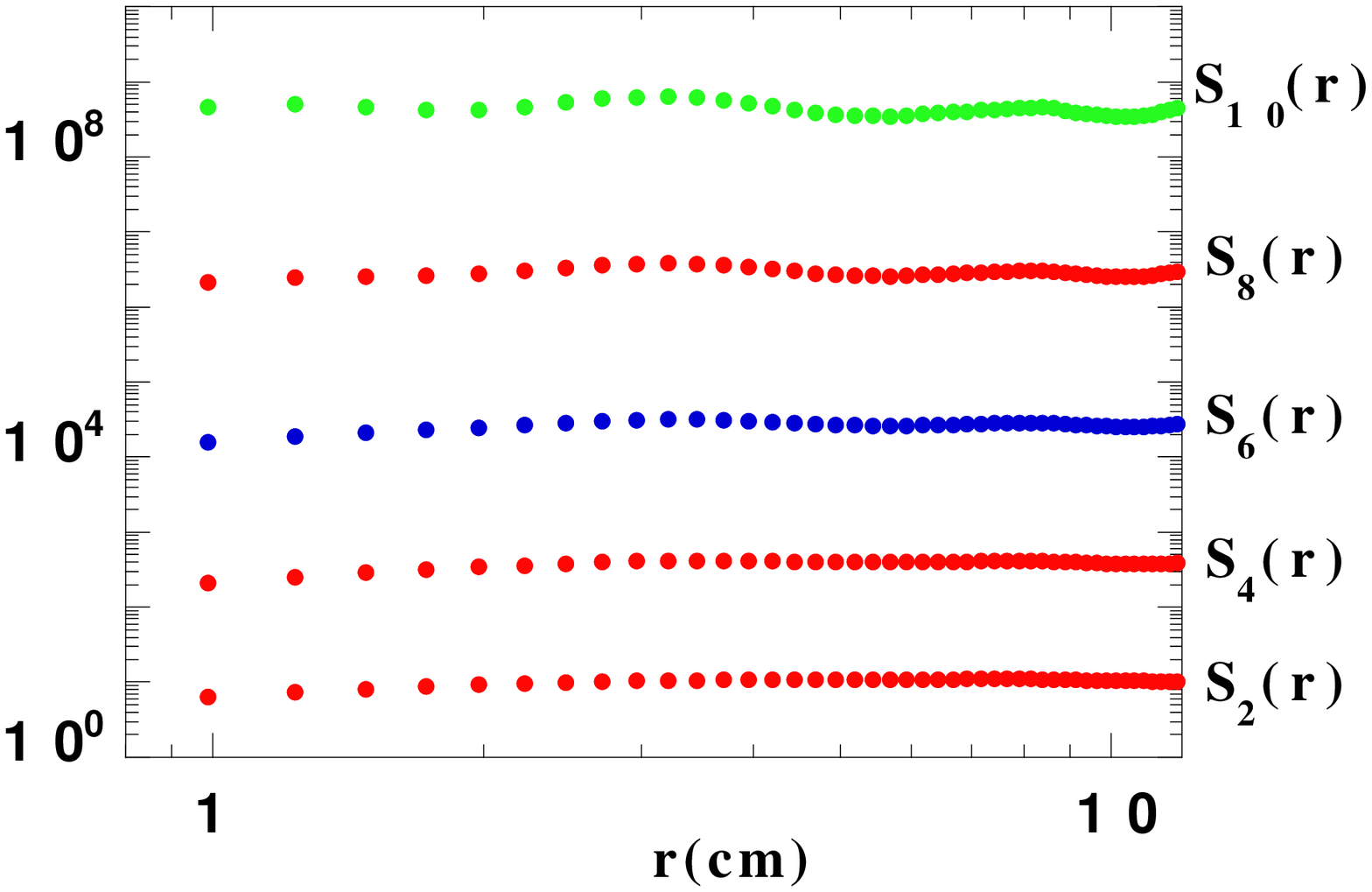}
}
\begin{figure}
\caption{Structure functions of the vorticity increments, of various 
orders between 2 and 10.}
\label{fstruc}
\end{figure}

\end{document}